\documentclass[pra,twocolumn,amscd,amsmath,amssymb,verbatim,letterpaper,showpacs]{revtex4}

\usepackage{graphicx} 
\usepackage{epstopdf} 
\usepackage{cancel} 

\newcommand{\leftsub}[2]{{\vphantom{#2}}_{#1}{#2}}
\begin{document}
\title{Two-particle quantum walks applied to the graph isomorphism problem}
\author{John King Gamble}
\email{jgamble@wisc.edu}
\author{Mark Friesen, Dong Zhou, Robert Joynt}
\author{S. N. Coppersmith}
\email{snc@physics.wisc.edu}
\affiliation{University of Wisconsin-Madison, Physics Department \\ 1150 University Ave, Madison, Wisconsin 53706, USA}

\date{\today}

\begin{abstract}
We show that the quantum dynamics of interacting
and noninteracting quantum particles are fundamentally
different in the context of solving a particular computational
problem.
Specifically, we consider the graph isomorphism problem, in which one wishes
to determine whether two graphs are isomorphic (related to each
other by a relabeling of the graph vertices), and focus on a class of
graphs with particularly high symmetry called strongly regular
graphs (SRG's).
We study the Green's functions that characterize the dynamical evolution
single-particle and two-particle quantum walks on pairs of non-isomorphic
SRG's
and show that interacting particles can distinguish non-isomorphic graphs
that noninteracting particles cannot.  
We obtain the following specific results:
(1) We prove that quantum walks of two noninteracting particles, 
Fermions or Bosons, cannot
distinguish certain pairs of non-isomorphic SRG's.
(2) We demonstrate
numerically that two interacting Bosons 
are more powerful than single particles and two
noninteracting particles, in that quantum walks of interacting bosons
distinguish all non-isomorphic pairs of SRGs that we examined.
By utilizing high-throughput computing to perform over 500 million direct comparisons between evolution operators, we checked all tabulated
pairs of non-isomorphic SRGs, including graphs with up to 64 vertices.
(3) By performing a short-time expansion of the evolution operator,
we derive distinguishing operators that provide analytic insight into the power of the interacting two-particle quantum walk.
\end{abstract}
\maketitle


\section{Introduction}\label{intro}
Random walks have been applied successfully to many problems in physics, as well as in many other disciplines, stretching from biology to economics \cite{Motwani:1996,Aleliunas:1979,Trautt:2006p2273,Sessions:1997p2221,Kilian200385}. 
Most of these applications use classical random walks (CRWs), in which quantum mechanical principles are not considered.
However, more recently, researchers have found that CRWs and quantum random walks (QRWs) can exhibit qualitatively different properties \cite{Aharonov:1993p2275,Bach2004562,Solenov:2006p2288}.   
 From a standpoint of algorithms research, these disparities lead to situations in which algorithms implemented by QRWs can be proven to run faster than the fastest possible classical algorithm \cite{Childs:2002p2294,Shenvi.PhysRevA.67.052307,Ambainis.2003,Ambainis.10.1109/FOCS.2004.54,Magniez.1250874.2007,Potocek.PhysRevA.79.012325,Reitzner.PhysRevA.79.012323}. 

 Besides being useful as theoretical models, simple QRWs have already been experimentally implemented in externally driven cavities \cite{Agarwal:2005p2283}, arrays of optical traps \cite{Eckert:2005p2286,Chandrashekar.PhysRevA.78.022314}, NMR systems \cite{Ryan:2005p2282}, and ion traps \cite{Travaglione:2002p2284,Schmitz.PhysRevLett.103.090504}. 
 This, coupled with new ideas for realistic physical implementations of non-trivial walks \cite{PhysRevA.80.060304}, indicates that studying algorithms cast as QRWs might lead to algorithms that are both powerful and experimentally viable.

Although QRWs are universal and therefore in principle can be used to implement any quantum algorithm \cite{Childs:2009p868}, in practice some information-theoretic problems lend themselves to a QRW approach more easily than others. 
Many of the problems that have been investigated are expressed naturally
using graphs, sets of vertices and edges, with each edge connecting two vertices. 
For example, the vertices of a graph might be taken to represent individuals, and the edges between them might indicate friendship.
A question one could ask is whether there is a subset of friends that are isolated from the rest of the group, which translates to the graph-theoretic question of whether the graph is disconnected.
Once the question to be answered is posed as a question about a graph, it is investigated by constructing a quantum Hamiltonian from that graph. 
The dynamics of the system is then analyzed using quantum mechanics, and is used to make statements about the original graph, hopefully giving insight to the answer of the original problem. 

This paper addresses the graph isomorphism (GI) problem, where, given two graphs, one must determine whether or not they are isomorphic (two graphs are isomorphic if one can be obtained from the other by a relabeling of the vertices). 
Although many special cases of GI have been shown to solvable in a time that scales as a polynomial of the number of vertices, the best general classical algorithm to date runs in time $\mathcal O \left( c^{ N^{1/2}\log N} \right)$, where $c$ is a constant and $N$ is the number of vertices in the graphs being compared \cite{Spielman:1996p1006}. 

GI has several properties analogous to those of factoring.  First, though it appears to be difficult, it is felt that it is unlikely to be NP-complete, since otherwise many complexity classes believed to be distinct would collapse \footnote{Specifically, it has been shown that graph isomorphism were NP-complete, the polynomial hierarchy collapses to level two \cite{Schoning:1988p2140}.}.
Second, both GI and factoring can be viewed as
hidden subgroup problems. 
In GI, one is looking for a hidden subgroup of the permutation group, while in factoring, one is looking for a hidden subgroup of the Abelian group.
The success of Shor's polynomial-time algorithm for factoring~\cite{Shor.1399018} has led several groups to investigate a hidden-subgroup approach to GI.  However, the obstacles facing such an approach have been shown to be formidable~\cite{Hallgren.1132603,Moore.1250868}.

Researchers have also recently attacked GI using various methods inspired by physical systems. 
Rudolph mapped the GI problem onto a system of hard-core atoms \cite{Rudolph:2002p1857}. 
One atom was used per vertex, and atoms $i$ and $j$ interacted if vertices $i$ and $j$ were connected by edges. He showed that pairs of non-isomorphic graphs exist whose original adjacency matrices have the same eigenvalues, while the induced adjacency matrices describing transitions between three-particle states have different eigenvalues. 
Gudkov and Nussinov proposed a physically-motivated classical algorithm to distinguish non-isomorphic graphs ~\cite{Gudkov:2002p1726}. 
Shiau \emph{et al.} proved that the simplest classical algorithm fails to distinguish some pairs of non-isomorphic graphs~\cite{Shiau:2005p1688} and also proved that continuous-time one-particle QRWs cannot distinguish some non-isomorphic graphs \cite{Shiau:2005p1688}. 
Douglas and Wang modified a  single-particle QRW by adding phase inhomogeneities, altering the evolution as the particle walked through the graph  \cite{Douglas:2008p1718}. 
Their approach was powerful enough to successfully distinguish many families of graphs considered to be difficult to distinguish, including all families of strongly regular graphs they tried. 
Most recently, Emms \emph{et al.} used discrete-time QRWs to build potential graph invariants \cite{Emms:2006p1925,Emms:2009p1930}. 
Through numerical spectral analysis, they found that these invariants could be used to distinguish many types of graphs by breaking  the eigenvalue degeneracies of many families of graphs that are difficult to distinguish.  

In addition to studying single-particle QRWs, Shiau et al. \cite{Shiau:2005p1688} performed numerical investigations of two-particle QRWs and presented evidence that interacting Bosons can distinguish non-isomorphic pairs that single-particle walks cannot. 
There, it was also found numerically that two-Boson QRWs with noninteracting particles do not distinguish some non-isomorphic pairs of graphs. 
In contrast to the approaches in  \cite{Douglas:2008p1718, Emms:2006p1925, Emms:2009p1930},
the two-particle QRW algorithm does not lower the symmetry of the system. 


In this paper, we extend the results in \cite{Shiau:2005p1688} on two-particle quantum walks in several ways.
First, we prove analytically that quantum walks of two non-interacting Bosons always fail to distinguish non-isomorphic pairs of strongly regular graphs (SRGs).
This result is surprising, since it has been shown in~\cite{Burioni:2000p2289,Buonsante:2002p1723,mancini-2007-918} that non-interacting Boson QRWs on graphs can give rise to effective statistical interactions, which significantly alter the dynamics of the system.
Second, we show that analysis of the behavior of non-interacting Fermions requires a more subtle treatment than was done in \cite{Shiau:2005p1688}; the result in \cite{Shiau:2005p1688} that some non-isomorphic SRGs could be distinguished by two noninteracting Fermions and not by two noninteracting Bosons arose because of an ambiguity in the choice of basis.
When the ambiguities involved with the basis choice for Fermions are accounted for, non-interacting Fermions fail to have any advantage over non-interacting Bosons. 
Third, we expand on the initial numerical results in \cite{Shiau:2005p1688}, exhaustively verifying, where only sampling was used before, that two-particle interacting Boson walks distinguish all the non-isomorphic pairs of SRGs with up to 64 vertices that have been tabulated.
To accomplish this, we used high-throughput computing techniques to perform over 500 million comparisons between evolution operators of graphs.
Finally, we examine the small-time expansion of the evolution operator, and use the two-particle interacting evolution to derive candidates for graph invariants, which appear in the fourth and sixth orders in time.

Our results demonstrate unambiguously that two-particle Bosonic quantum walks have more computational power if the particles are interacting, because interacting walks can be used to distinguish non-isomorphic graphs that noninteracting particles cannot.

The paper is organized as follows.  
Section~\ref{sec:definitions} introduces relevant background and definitions to QRWs on graphs, including a brief overview of the strongly regular graphs (SRGs) on which the algorithms are tested and also a review of the one-particle QRW algorithm considered in \cite{Shiau:2005p1688}.
Section~\ref{secproof} proves that QRWs of two non-interacting Bosons do not distinguish non-isomorphic SRGs with the same family parameters.
Section~\ref{secproofF} analyzes the QRW of two non-interacting Fermions.
It shows that improper basis choice can lead to false-distinguishing, and that when the basis is chosen consistently, QRWs with two noninteracting fermions are unable to distinguish some pairs of non-isomorphic SRGs.
Section~\ref{sec:numerics} shows through exhaustive simulation that all tabulated families of SRGs are successfully distinguished by a two hard-core Boson QRW.
In section~\ref{sec:smalltime} a short-time expansion of the evolution operator is computed, and distinguishing operators present in the two hard-core Boson QRW are identified.
Finally, Section~\ref{sec:discussion} summarizes and discusses the possible implications of our results for the development of algorithms based on interacting QRWs for distinguishing non-isomorphic graphs.

\section{Background \& Definitions}
This section describes the background and definitions necessary to study QRWs on graphs. First, we introduce the graph-theoretic concepts we will need, including the notions of of the adjacency matrix and the spectrum of a graph. Then, we consider how to use these tools to construct a physical process through the definition of Hamiltonian and Green's functions, for both one and two particles. Next, we detail the relevant properties of SRGs. Finally, we review the method used in \cite{Shiau:2005p1688} to use the properties of SRGs to show that the Green's functions of single-particle QRWs do not distinguish non-isomorphic SRGs with the same family parameters.
\label{sec:definitions}

\subsection{Constructing walks on graphs}
In this section, we describe how to form QRWs on graphs. 
A graph $G = (V,E)$ is a set vertices $V$ and edges $E$.
The vertices are usually labeled by integer indices, and the edges are unordered pairs of vertices.
Two vertices that share an edge are called \emph{connected}, while two vertices that do not are called \emph{disconnected}.
The total number of neighboring vertices of a particular vertex is called its \emph{degree}.
For example, the graph $G = \left( \{1,2,3,4\},\{(1,2),(2,3),(3,4),(4,1)\} \right)$ is a cycle graph with four vertices.
Two graphs are \emph{isomorphic} if they can be made identical by relabeling their vertices.
For example, the graph $H= \left( \{1,2,3,4\},\{(1,3),(3,2),(2,4),(4,1)\} \right)$ is isomorphic to $G$, since after relabeling $2 \leftrightarrow 3$, the two graphs are equivalent.

Graphs are conveniently expressed algebraically as \emph{adjacency matrices}. An adjacency matrix $A$ of a graph with $N$ vertices is an $N \times N$ matrix in the basis of vertex labels, with $A_{ij}=1$ if vertices $i$ and $j$ are connected by an edge, and zero otherwise. The adjacency matrix for the graph $G$ is
\begin{equation}
\mathbf A_G = \left(
\begin{array}{cccc}
 0 & 1 & 0 & 1 \\
 1 & 0 & 1 & 0 \\
 0 & 1 & 0 & 1 \\
 1 & 0 & 1 & 0
\end{array}
\right).
\end{equation}
The \emph{spectrum} of a graph is the eigenvalue spectrum of its adjacency matrix. The spectrum of $G$ is $\{ -2 , 0 ,2 \}$, with $0$ two-fold degenerate.

To form a QRW on a graph, we first define a Hamiltonian. We will use the Hubbard model, with each site corresponding to a graph vertex. A particle can make a transition between two sites if the associated vertices are connected. In addition, if two walkers happen to simultaneously occupy a site, we impose a double-occupation energy cost $U$. Our Hamiltonian is 
\begin{equation}\label{ham_gen}
\mathbf H = - \sum_{i,j} A_{ij} c^{\dagger}_i c_j +\frac{U}{2} \sum_i \left( c^{\dagger}_ic_i \right)\left( c^{\dagger}_ic_i -1\right),
\end{equation}
where $c$ and $c^{\dagger}$ are Boson or (spinless) Fermion creation and annihilation operators. 
If we restrict ourselves to single-particle states, we find matrix elements
\begin{equation}
\left< i \right| \mathbf H \left| j \right> = - A_{ij}.
\end{equation}
Hence, we can easily identify the a single-particle Hamiltonian
\begin{equation}\label{opform}
\mathbf H_{1P} = - \mathbf A.
\end{equation}
Similarly, we can define two-particle Hamiltonians by their matrix elements. 
For these, we need to use either Bosonic or Fermionic basis states.
The Boson  states are
\begin{equation}\label{defbosonbasis}
\left| i j \right>_B \equiv  \left\{
     \begin{array}{lr}
       \frac{ 1}{\sqrt{2}} \left(\left| i j \right> +  \left| j i \right> \right)& : \, i \neq j \\
       \left| i  i  \right> & \, : i = j
     \end{array}
   \right.
\end{equation}
and the Fermion states are
\begin{equation}
\left| i j \right>_F \equiv  \frac{ 1}{\sqrt{2}} \left(\left| i j \right> -  \left| j i \right> \right).
\end{equation}
We now restrict ourselves to two particles, where we define $\mathbf H_{2B}$ and $\mathbf H_{2F}$ to be the two-particle Boson and Fermion Hamiltonians. These are special cases of eq. \ref{ham_gen}, with matrix elements
\begin{widetext}
\begin{eqnarray} \label{bosonelements}
\leftsub{B}{\left< ij \right|} \mathbf H_{2B }\left| kl \right>_B  &\equiv& \leftsub{B}{\left< ij \right|} \mathbf H \left| kl \right>_B =
 \left\{
     \begin{array}{lr}
      - \delta_{ik} A_{jl} - \delta_{jl} A_{ik} - \delta_{il} A_{jk} - \delta_{jk} A_{il} & : \,\, i \neq j \,\, \mathrm{and} \,\,  k \neq l \\
       U \delta_{ik}& : \, i = j\,\, \mathrm{and} \,\,   k = l \\
       \frac{-1}{\sqrt{2}} \left(  \delta_{ik} A_{jl}+ \delta_{jl} A_{ik}+ \delta_{il} A_{jk} + \delta_{jk} A_{il}  \right) & :  \,\, i = j \,\, \mathrm{xor} \,\,  k = l
     \end{array},
   \right. \nonumber \\
\leftsub{F}{\left< i  j \right|} \mathbf H_{2F} \left| k l \right>_F  &\equiv&\leftsub{F}{\left< ij \right|} \mathbf H \left| kl \right>_F  =
 \begin{array}{lr}
 A_{ik} \delta_{jl} + A_{jl} \delta_{ik}   - A_{il} \delta_{jk} - A_{jk} \delta_{il} & : \,\, i \neq j \,\, \mathrm{and} \,\,  k \neq l ,
\end{array}
\end{eqnarray}
\end{widetext}
respectively, where the matrix elements are found directly from Eq. \ref{ham_gen} through the application of appropriate commutation relations.

From each of these Hamiltonians, we define the QRW time-evolution operator as
\begin{equation}\label{1pevolution}
\mathbf U= e^{-i t \mathbf H}, 
\end{equation}
where we set $\hbar=1$ for notational convenience. 

To study the dynamics of the system, we define the two-particle Green's function (GF), which relates the wavefunctions at a time $t$ to those at time $t=0$. 
For two-particle position states $\left| \psi \right>$ and $\left| \psi ' \right>$, it is
\begin{equation}\label{gfdef}
\mathcal  G \left( \psi(0) , \psi'(t) \right) =  \left< \psi(0) \Big | \psi '(t) \right> = \left< \psi(0) \right| \mathbf U \left| \psi '(0) \right>.
\end{equation}
Letting $(\psi, \psi')$ run over a complete two-particle basis, $\mathcal G(\psi(0),\psi'(t))$
considered at a fixed time provides us with a set of $N^2(N+1)^2/4$ complex numbers for Bosons or $N^2(N-1)^2/4$ complex numbers for Fermions.
These lists completely characterizes the dynamics of the system.  Hence, when we analyze QRWs, we say that a
particular scheme \emph{distinguishes} two graphs if their Green's functions supply us with two different lists.

\subsection{Strongly Regular Graphs}\label{subsec:strongly_regular_graphs}
Our major results in this paper focus on algorithms over strongly regular graphs (SRGs), which are difficult to distinguish.
In this section, we develop the properties of SRGs we will need for our later analysis. 

A SRG is a graph in which (a) all vertices have the same degree, (b) each pair of neighboring vertices has the same number of shared neighbors, and (c) each pair of non-neighboring vertices has the same number of shared neighbors. This definition permits SRGs to be categorized into families by four integers $(N,k,\lambda, \mu)$, each of which might contain many non-isomorphic members.
Here, $N$ is the number of vertices in each graph, $k$ is the degree of each vertex (k-regularity), $\lambda$ is the number of of common neighbors shared by each pair of adjacent vertices, and $\mu$ is the number common neighbors shared by each pair of non-adjacent vertices. 

\begin{figure}[tb] 
\begin{center} 
\includegraphics[width=1.0 \linewidth]{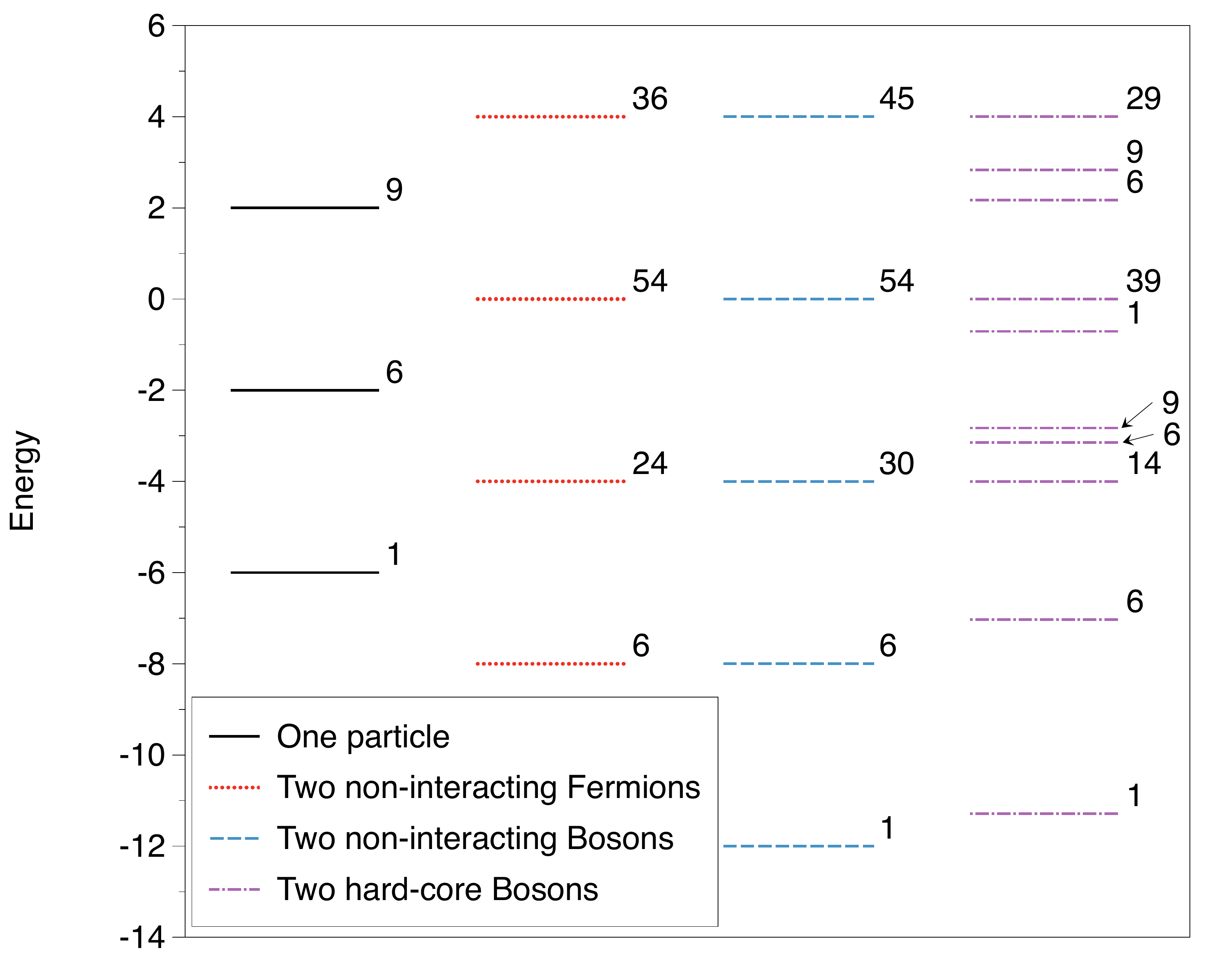}
\end{center} 
\caption{The energy spectra for several types of QRWs on the SRG family (16,6,2,2), where the parameters $\left( N, k, \lambda, \mu \right)$ define an individual SRG family.
$N$ is the number of vertices in each graph, $k$ is the degree of each vertex (k-regularity), $\lambda$ is the number of of common neighbors shared by each pair of adjacent vertices, and $\mu$ is the number common neighbors shared by each pair of non-adjacent vertices.  Both graphs in the family have the same spectra.
In all four panels, the Hamiltonian used is the Hubbard model (eq. \ref{ham_gen}), with $U=0$ in the noninteracting cases and $U \rightarrow \infty$ for the hard-core Bosons.
The degeneracies are given to the right of each level.
\label{fig:SRGSpectra} }
\end{figure}

Using the stringent constraints placed on SRGs, one can show that, regardless of size, the spectrum of any SRG only has three distinct values \cite{godsil:2001}:
\begin{equation}
\lambda_0 = -k,
\end{equation}
which is non-degenerate, and
\begin{equation}
\lambda_{1,2} = -\frac{1}{2} \left( \lambda - \mu \pm \sqrt{N} \right),
\end{equation}
which are both highly degenerate. 
Both the value and degeneracy of these eigenvalues depend only on the family parameters, so within a particular SRG family, all graphs are cospectral \cite{godsil:2001}. 
Further, the spectra of the two-particle Hamiltonians formed from SRG adjacency matrices, as described in Eq. \ref{bosonelements}, are also highly degenerate.
As shown in figure \ref{fig:SRGSpectra} for the family (16,6,2,2), the interacting case gives us the largest number of distinct energy levels.
These highly degenerate spectra are one reason why distinguishing non-isomorphic SRGs is difficult---it is known that distinguishing non-isomorphic graphs with spectra with bounded degeneracy can be done with polynomially bounded resources~\cite{luks1982}.

The adjacency matrix of any SRG satisfies the useful relation \cite{godsil:2001}:
\begin{equation} \label{SRGAlgebra}
\mathbf A^2 = (k-\mu) \mathbf I + \mu \mathbf J + (\lambda - \mu) \mathbf A,
\end{equation}
where $\mathbf I$ is the identity and $\mathbf J$ is the matrix of all ones
($J_{ij}=1$ for all $i,j$). 
Since $\mathbf J^2 = N \mathbf J$ and $\mathbf I^2 = \mathbf I$, this forms a three-dimensional algebra, and we can write
\begin{equation} \label{SRGAlgebra2}
\mathbf A^n = \alpha_n \mathbf I + \beta_n \mathbf J + \gamma_n \mathbf A,
\end{equation}
where $\alpha$, $\beta$, and $\gamma$ are functions only of the family parameters. 
That is, all SRGs of the same family have the same coefficients. 

Although many SRGs are known \cite{srglistweisstein}, there are substantially fewer tabulated families with more than one non-isomorphic member.
Complete and partial families of SRGs have been tabulated through combinatorial techniques \cite{spencesrgs,mckaysrgs}, 
which we use in sec. \ref{sec:numerics} to perform numerical tests of our algorithms.

\subsection{Review of one-particle algorithm}
In this section, we review the method used in Ref. \cite{Shiau:2005p1688} to prove that a single-particle QRW cannot distinguish non-isomorphic members of SRG families. Formally, this means that we must show that any two SRGs of the same family parameters have the same single-particle Green's functions $\mathcal  G_{1P}$.

First, we consider the adjacency matrix $\mathbf A$, and suppose it belongs to the strongly regular graph family $(N,k,\lambda, \mu)$.
Then, by Eq. \ref{opform}, we know that the single-particle hamiltonian is $\mathbf H_{1P} = - \mathbf A$, and so by Eq. \ref{1pevolution}, the QRW evolution operator is $\mathbf U_{1P} = e^{i t \mathbf A}$. 
But since $\mathbf A$ is a SRG, we make use of the algebra defined in Eq. \ref{SRGAlgebra2} to write 
\begin{equation}\label{eqn:defcoeffs}
\mathbf U_{1P} = \alpha \mathbf I + \beta \mathbf J + \gamma \mathbf A, 
\end{equation}
where the coefficients depend only on the family parameters. 

Following Shiau et al., we investigate the relevant Green's functions, $\mathcal  G_{1P}(i,j) = \left< i \right| \mathbf U_{1P} \left| j \right>$.
We first consider the diagonal elements, each of which contains a contribution of $\alpha$ from $\mathbf I$ and $\beta$ from $\mathbf J$. 
Note that there is no contribution from $\mathbf A$, because it is entirely off-diagonal. 
Hence, the $N$ diagonal Green's functions are all equal to $\alpha + \beta$. 
For the off-diagonal elements, $\mathbf I$ never contributes and $\mathbf J$ always does. 
However, $\mathbf A$ contributes to only some of the elements. 
More precisely, each column of $\mathbf A$ contains exactly $k$ ones (entirely in the off-diagonal), since each vertex is of degree $k$. 
Hence, there are $kN$ off-diagonal Green's functions of the form $\beta + \gamma$, and the
remaining $N^2-N-kN$ are equal to $\beta$. 

As can be seen from Eq. \ref{SRGAlgebra2}, $\alpha$, $\beta$, and $\gamma$ all depend only on the family parameters.  Therefore, the one-particle evolution for any graph in the same family will have the same GFs, and the algorithm based on single-particle quantum evolution fails to distinguish any non-isomorphic SRGs that are in the same family.

\section{Proof that QRWs with two non-interacting Bosons do not distinguish non-isomorphic SRGs in the same family}\label{secproof}

\begin{table*}[tb]
\caption{This table enumerates the Green's functions for the QRW with two non-interacting Bosons on a SRG with family parameters $(N,k,\mu,\lambda)$.
The Hamiltonian considered is $ \mathbf H_B = -1/2 \cdot \left( \mathbf I + \mathbf S \right) \left( \mathbf A \oplus \mathbf A \right)$, where $\mathbf S$ swaps the two particles and $\mathbf A$ is the adjacency matrix of the graph. 
Because the $\mathbf A$, $\mathbf I$, and $\mathbf J$ form an algebra, the parameters $\alpha$, $\beta$, and $\gamma$ are the same for every graph in an SRG family (Eq.~(\ref{SRGAlgebra2}).
The evolution operator for noninteracting Bosons, $\mathbf U_B = 1/2  \left( \mathbf I + \mathbf S \right)\mathbf U^{1P} \otimes  \mathbf U^{1P}$, contains terms bilinear in $\mathbf I$, $\mathbf J$, and $\mathbf A$, with coefficients that can be written in terms of $\alpha$, $\beta$, and $\gamma$. 
The GFs, formed by taking matrix elements of $\mathbf U_B$ (Eq.~(\ref{gfdef}), are divided into symmetry classes $(a,b)$, where $a$ indicates the number of distinct basis indices and $b$ is the number of indices shared between the left and right sides.
For example, $ \left< 3 4 \right|_B U_B \left| 2 4 \right>_B$,
 where $\left| ij  \right>_B $ indicates identical Bosons on vertices $i$ and $j$, falls into the symmetry class $(3,1)$, since it has three distinct indices (2,3,4) and the left and right side have one index in common (2). 
Since the total number of entries with any given particular value can be counted in terms of numbers that are constant for a given set of family parameters, the GFs generated by non-isomorphic members of the same SRG family must have the same values and the same degeneracies.
 \label{NIBosontable}
}
\begin{tabular}{| c | l | l |}
\hline 
Element Class & Value of Element & Number of Occurrences \\
\hline
\hline
(4,0) & $4 \beta \gamma + 2 \gamma^2+2\beta^2$ 									& $1/4 \cdot N \left(k^2 (\mu +1)+k \left(\lambda ^2-\lambda  (\mu +2)+\mu -1\right)-2 (N-1) \mu \right)$ \\
\hline
 	& $3 \beta \gamma + \gamma^2 + 2 \beta ^2$ 									& $N \mu  (N-k-1) (k+\lambda -\mu )$ \\
 \hline
 	& $2 \beta \gamma + 2 \beta^2 + \gamma^2$ 									& $1/(2 k) \cdot \left[ N (N-k-1) \left(k^3-2 k^2 \mu +(N-1) \mu ^2\right)\right]$ \\
\hline
 	& $2 \beta \gamma + 2 \beta^2 $ 											& $1/k \cdot \left[N (N-k-1) \left(k^3-k^2 (2 \mu +1)+(N-1) \mu ^2\right)\right]$ \\
\hline
 	& $ \beta \gamma + 2 \beta^2 $ 											& $1/k \cdot \left[N (k-N+1) (k-\mu ) (k (2 k-N+2)-N \mu +\mu )\right]$ \\
\hline
 	& $2 \beta^2 + \gamma^2$ 												& $1/(4 k) \cdot \big[N (N-k-1) \big(k (-3 k N+k (3 k+8)$ \\
 	& 																	& $+N^2-5 N+6)-2 k (k+1) \mu +(N-1) \mu ^2\big)\big]$ \\
\hline
	& Subtotal:															&$1/4 \cdot N(N-1)(N-2)(N-3)$ \\

\hline
\hline

(3,0) & $4/\sqrt{2} \cdot  \beta \gamma + 2/\sqrt{2} \cdot \beta^2+2/\sqrt{2} \cdot \gamma^2$ 		& $k N (k-\lambda -1)+k N \lambda$ \\
\hline
	& $2/\sqrt{2} \cdot  \beta \gamma + 2/\sqrt{2} \cdot \beta^2$ 							& $2 k N (N-k-1)$ \\
\hline
	& $2/\sqrt{2} \cdot \beta^2$											 		& $N (k-N+1) (k-N+2)$ \\
\hline
	& Subtotal:																&$N(N-1)(N-2)$ \\
	
\hline
\hline

(3,1) & $\alpha \beta + \alpha \gamma + 3 \beta \gamma + 2 \beta^2 + \gamma^2 $		 		& $k N \lambda$ \\
\hline
	& $\alpha \beta + 2 \beta \gamma + 2 \beta^2 + \gamma^2$ 							& $N (N -1 - k ) \mu $ \\
\hline
	& $\alpha \beta +\alpha \gamma+ 2 \beta \gamma + 2 \beta^2$						& $2N (N -1 - k ) \mu$ \\
\hline
	& $\alpha \beta  +\alpha \gamma+  \beta \gamma + 2 \beta^2$							& $k N (-2 k + N + \lambda)$ \\
\hline
	& $\alpha \beta + \beta \gamma + 2 \beta^2$								 		& $2k N (-2 k + N + \lambda)$ \\
\hline
	& $\alpha \beta +  2 \beta^2$								 					& $N(1 + k - N) (2 + 2 k - N - \mu)$ \\
\hline
	& Subtotal:											2					&$N(N-1)(N-2)$ \\

\hline
\hline

(2,0) & $2 \beta \gamma + \beta^2 + \gamma^2 $		 								& $k N$ \\
\hline
	& $ \beta^2 $		 														& $N(N-k-1)$ \\
\hline
	& Subtotal:																&$N(N-1)$ \\

\hline
\hline

(2,1)	& $2/\sqrt{2} \cdot \left( \alpha \beta + \alpha \gamma + \beta \gamma + \beta^2 \right)$		& $2 k N$ \\
\hline
	& $2/\sqrt{2} \cdot \left( \alpha \beta + \beta^2 \right)$								& $2 N (N-k-1) $\\
\hline
	& Subtotal:																&$2N(N-1)$ \\

\hline
\hline

(2,2) & $\alpha^2 + 2 \alpha \beta + 2 \beta \gamma + 2 \beta^2 + \gamma^2$					& $1/2 \cdot k N $ \\
\hline
	& $\alpha^2 + 2 \alpha \beta + 2 \beta^2$											& $1/2 \cdot N (N-k-1)$ \\
\hline
	& Subtotal:																&$1/2 \cdot N(N-1)$ \\

\hline
\hline
(1,2)	& $\alpha^2 + 2 \alpha \beta + \beta^2$											& N \\

\hline
\hline
Total: &																		&$1/4 \cdot N^2(N+1)^2$ \\
\hline
	
\end{tabular}
\end{table*}
 
We now develop exact expressions for the time evolution operators for two non-interacting Bosons, and subsequently show that this evolution cannot be used to distinguish non-isomorphic SRGs in the same family.
Although this result may seem expected, recent efforts \cite{mancini-2007-918,Burioni:2000p2289,Buonsante:2002p1723} have
demonstrated that non-interacting QRWs on non-translationally invariant graphs lead to effective, statistical interactions, resulting in 
rich physical phenomena such as Bose-Einstein condensation.
The method used here is analogous to that used in Ref.~\cite{Shiau:2005p1688} to show that one-particle QRWs cannot distinguish non-isomorphic SRGs from the same family, but with a more complex implementation.

First, we note that we may write the Hamiltonian for any two-Boson QRW as
\begin{equation}\label{bosonoperatorham}
\mathbf H_{2B} = -\frac{1}{2}\left( \mathbf I + \mathbf S \right) \left( \mathbf A \oplus \mathbf A \right)+ U \mathbf R,
\end{equation}
where $\mathbf A \oplus \mathbf A = \mathbf A \otimes \mathbf I + \mathbf I \otimes \mathbf A$ is a Kronecker sum, the matrix special case of a direct sum, and
\begin{equation}
\mathbf S = \sum_{i,j} \left| i j \right>\left< j i \right|, \, \, \mathbf R  = \sum_{i} \left| i i \right> \left< i i \right|.
\end{equation}
The demonstration that Eq. \ref{bosonoperatorham} is equivalent to Eq. \ref{bosonelements} is given in appendix \ref{twoptclcheck}.
For noninteracting Bosons $U=0$, and the
evolution operator is
\begin{eqnarray}
\mathbf U_{2B} 
&=& e^{- i t \mathbf H_{2B}} \nonumber \\&=&  \sum_{n=0}^\infty \frac{1}{n!} \left(i t \frac{1}{2} \left( \mathbf I + \mathbf S \right) \left(\mathbf A \oplus \mathbf A \right) \right)^n,
\end{eqnarray}
where $\mathbf U_{2B}$ is shorthand to refer only to the non-interacting case.
Now, by the definitions of the Kronecker sum and $\mathbf S$, it is easy to show that $\left[  \left( \mathbf I + \mathbf S \right) , \left(\mathbf A \oplus \mathbf A \right)  \right] = \mathbf 0$. 
Hence,
\begin{equation}
\mathbf U_{2B} = \sum_{n=0}^\infty \frac{1}{n!} \left(i t \frac{1}{2}\right)^n \left( \mathbf I + \mathbf S \right)^n \left(\mathbf A \oplus \mathbf A \right)^n.
\end{equation}
But note that $\mathbf S^2 = \mathbf I$, so $\left(\mathbf I + \mathbf S \right)^n = 2^{n-1} \left( \mathbf I + \mathbf S \right)$. It follows that
\begin{eqnarray}
\mathbf U_{2B} &=& \frac{1}{2} \left( \mathbf I + \mathbf S \right) e^{it \mathbf A \oplus it \mathbf A}.
\end{eqnarray}

Since each matrix $A$ in the Kronecker (direct) sum exponentiates in its own product space, we can write $e^{\mathbf A \oplus \mathbf B} = e^{\mathbf A} \otimes e^{\mathbf B}$, which leads to
\begin{eqnarray}
\mathbf U_{2B} &=& \frac{1}{2} \left( \mathbf I + \mathbf S \right) e^{it \mathbf A} \otimes e^{it \mathbf A} \nonumber \\
&=& \frac{1}{2} \left( \mathbf I + \mathbf S \right) \mathbf U^{1P} \otimes  \mathbf U^{1P} ,
\end{eqnarray}
where the single-particle evolution operator $\mathbf U_{1P}$ is defined in Eq. \ref{1pevolution} for $\mathbf H =- \mathbf A$. Since the Boson states are symmetric under particle interchange, we have matrix elements
\begin{equation}
\leftsub{B}{\left< i j \right|} \mathbf U_{2B} \left| kl \right>_B = \leftsub{B}{\left< i j \right|} \mathbf U_{1P} \otimes  \mathbf U_{1P} \left| kl \right>_B .
\end{equation}
Expanding this using eq. \ref{eqn:defcoeffs}, we have
\begin{eqnarray}\label{U2Belements}
\leftsub{B}{\left< i j \right|} \mathbf U_{2B} \left| kl \right>_B &=& \leftsub{B}{\left< i j \right|} \Big( \alpha^2 \mathbf I \otimes \mathbf I + \beta^2 \mathbf J \otimes \mathbf J \nonumber \\
&+& \gamma^2 \mathbf A \otimes \mathbf A  + \alpha \beta \left( \mathbf J \oplus \mathbf J \right)+ \alpha \gamma \left( \mathbf A \oplus \mathbf A \right)  \nonumber \\
&+& \beta \gamma \left( \mathbf J \otimes \mathbf A + \mathbf A \otimes \mathbf J \right) \Big) \left| kl \right>_B.
\end{eqnarray}

Now that we have determined the matrix elements of $ \mathbf U_{2B}$, we can work out all the cases for eq. \ref{U2Belements}, which are the GFs of the system.  We find that there are 22 possible values, each of which can be written as an explicit function of $\alpha$, $\beta$, and $\gamma$. 
Since the values of the matrix elements are all only functions of SRG family parameters, they are the same for all graphs in the same SRG family.
To show that the GFs are the same across a SRG family, we also need to show that the number of occurrences of each value is also a function only of SRG parameters.
In Appendix \ref{countingboson} we count all the types of these GFs in terms of SRG family parameters, with the results shown in table \ref{NIBosontable}. 

Because we have shown that both the values and number of occurrences of all of the two-particle Green's functions for noninteracting Bosons can be written in terms of the family parameters $N$, $k$, $\lambda$, and $\mu$, we demonstrated that two noninteracting Bosons cannot distinguish non-isomorphic SRGs of the same family.

\section{Analysis of the noninteracting two-Fermion evolution for SRGs}\label{secproofF}
\begin{table*}[tb]
\caption{Enumeration of the Green's functions for QRWs of two non-interacting Fermions. 
The Hamiltonian for two noninteracting Fermions is $ \mathbf H_{2F} = 1/2 \cdot \left( \mathbf I - \mathbf S \right) \left( \mathbf A \oplus \mathbf A \right)$, where $\mathbf S$ swaps the two particles and $\mathbf A$ is the adjacency matrix of the graph. 
Hence, the two-Fermion evolution operator, defined with $t \rightarrow -t$ to keep $U^{1P}$ running with forward time, is $\mathbf U_F = 1/2  \left( \mathbf I - \mathbf S \right)\mathbf U^{1P} \otimes  \mathbf U^{1P}$. 
It contains terms bilinear in $\mathbf I$, $\mathbf J$, and $\mathbf A$, with coefficients written as combinations of $\alpha$, $\beta$, and $\gamma$, where
the parameters $\alpha$, $\beta$, and $\gamma$, are defined in Eq.~(\ref{eqn:defcoeffs}).
We divide the matrix elements $\left< i j \right|_F U_{2F} \left| k l \right>_F$ of the GFs into classes $(a,b)$, where $a$ indicates the number of distinct indices and $b$ is the number of indices shared between the left and right sides.
For example, $\left< 3 4 \right|_F U_F \left| 2 4 \right>_F$ falls into class $(3,1)$, since it has the three distinct indices (2,3,4) and the left and right side have the index (2) in common. 
The $\pm$ next to some of the element values indicates that the count applies to all elements with the given magnitude; this is done because the number of elements with each sign depends on the choice of two-particle basis.
The total number of matrix elements with a given absolute value can be expressed in terms of the SRG family parameters.  Therefore, we conclude that the matrix elements of the GFs of two non-isomorphic members of the same SRG family must be equivalent up to sign differences. \label{NIFermiontable}
}
\begin{tabular}{| c | l | l |}
\hline 
Element Class & Value of Element & Number of Occurrences \\
\hline
\hline
(4,0) & $0$						 										& $1/(4 k ) \cdot n \big(-6 k^4+2 k^3 (5 n+6 \mu -7)-4 k^2 (n-1) (n+3 \mu -2)$ \\
 	& 																	& $+k (n-1) \left((n-5) n-6 \mu ^2+6\right)+6 (n-1)^2 \mu ^2\big)$ \\
\hline
 	& $\pm \left( \beta \gamma + \gamma^2 \right)$ 								& $N \mu \cdot  (N-k-1) (k+\lambda -\mu )$ \\
\hline	
 	& $   \pm \left( \gamma^2+2 \beta \gamma \right)$ 								& $1/k \cdot \left[ N (N-k-1) \left(k^3-2 k^2 \mu +(N-1) \mu ^2\right)\right]$ \\
\hline
 	& $ \pm \beta \gamma $ 													& $1/k \cdot \left[N (k-N+1) (k-\mu ) (k (2 k-N+2)-N \mu +\mu )\right]$ \\
\hline
	& Subtotal:															&$1/4 \cdot N(N-1)(N-2)(N-3)$ \\

\hline
\hline

(3,1) & $\pm \left( \alpha \beta + \alpha \gamma -\beta \gamma - \gamma^2 \right)$		 	& $k N \lambda$ \\
\hline
	& $\pm \left( \alpha \beta -2\beta \gamma - \gamma^2 \right)$ 						& $N (N -1 - k ) \mu $ \\
\hline
	& $\pm \left( \alpha \beta +\alpha \gamma \right)$								& $2N (N -1 - k ) \mu$ \\
\hline
	& $\pm \left( \alpha \beta +\alpha \gamma + \beta \gamma \right)$					& $k N (-2 k + N + \lambda)$ \\
\hline
	& $\pm \left( \alpha \beta -\beta \gamma \right)$								& $2k N (-2 k + N + \lambda)$ \\
\hline
	& $\pm \alpha \beta $								 					& $N(1 + k - N) (2 + 2 k - N - \mu)$ \\
\hline
	& Subtotal:															&$N(N-1)(N-2)$ \\

\hline
\hline

(2,2) & $\alpha^2 + 2 \alpha \beta - 2 \beta \gamma - \gamma^2$						& $1/2 \cdot k N $ \\
\hline
	& $\alpha^2 + 2 \alpha \beta $												& $ 1/2 \cdot N (N-k-1)$ \\
\hline
	& Subtotal:															&$1/2 \cdot N(N-1)$ \\

\hline
\hline
Total: &																	&$1/4 \cdot N^2(N-1)^2$ \\
\hline
	
\end{tabular}
\end{table*}

In this subsection we consider the analogous evolution generated by two non-interacting Fermions.
This analysis is more complicated than for Bosons because 
changing the two-particle basis can introduce sign changes.  
If this sign ambiguity is not accounted for properly, the algorithm may falsely distinguish two graphs that are actually isomorphic.  

The Hamiltonian for the two-Fermion QRW, $\mathbf H_{2F}$, is
\begin{equation}\label{fermionoperatorham}
\mathbf H_{2F} = \frac{1}{2}\left( \mathbf I - \mathbf S \right) \left( \mathbf A \oplus \mathbf A \right),
\end{equation}
where, again, $I$ is the identity, $S$ is the operator that swaps the two particles, $A$ is the adjacency matrix of the graph, and $\oplus$ denotes a direct sum.
We follow the same logic we did for the Bosons but let $t \rightarrow -t$,
\begin{eqnarray}
\mathbf U_{2F}(-t) \equiv \mathbf U_{2F}~.
\end{eqnarray}
This way, the single-particle evolution operator $\mathbf U_{1P}$ still has time running forwards.
The matrix elements of $\mathbf U_{2F}$ are given by
\begin{eqnarray}
\leftsub{F}{\left< i j \right|} \mathbf U_{2F}\left| kl \right>_F &=& \leftsub{F}{\left< i j \right|} \Big( \alpha^2 \mathbf I \otimes \mathbf I + \beta^2 \mathbf J \otimes \mathbf J + \gamma^2 \mathbf A \otimes \mathbf A \nonumber \\
&+& \alpha \beta \left( \mathbf J \oplus \mathbf J \right)+ \alpha \gamma \left( \mathbf A \oplus \mathbf A \right) \nonumber \\
&+& \beta \gamma \left( \mathbf J \otimes \mathbf A + \mathbf A \otimes \mathbf J \right) \Big) \left| kl \right>_F.
\end{eqnarray}

\begin{figure}[tb] 
\begin{center} 
\includegraphics[width=0.7 \linewidth]{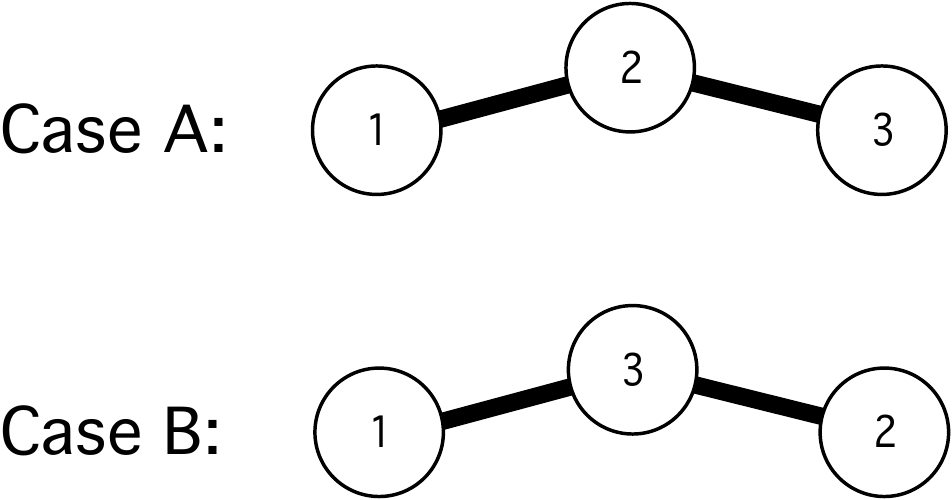}
\end{center} 
\caption{Two clearly isomorphic graphs. 
Graph $A$ differs from graph $B$ only by the labeling of vertices 2 and 3. 
Despite this, some matrix elements of two-particle Fermion evolution operators $\mathbf U_F$ of the two graphs have opposite signs.   This result implies that using the numerical values of these matrix elements produces a false-positive: two isomorphic graphs are falsely distinguished.\label{fig:fermprob} }
\end{figure}

We now show that there are sign ambiguities in $\mathbf U_F$ that arise from the choice of basis states that one uses
when converting a graph to an adjacency matrix. 
As an example, consider the two isomorphic graphs shown in figure \ref{fig:fermprob}, which have adjacency matrices
\begin{equation}
\mathbf A = \left(
\begin{array}{ccc}
 0 & 1 & 0 \\
 1 & 0 & 1 \\
 0 & 1 & 0
\end{array}
\right), \, \, \, \mathbf B = \left(
\begin{array}{ccc}
 0 & 1 & 1 \\
 1 & 0 & 0 \\
 1 & 0 & 0
\end{array}
\right).
\end{equation}
We wish to write down the two-particle Hamiltonians using eq. \ref{bosonelements}.
However, we must first pick a basis. 
That is, for each pair of sites $\left| i j \right> \equiv \left| j i \right>$, we are free to pick either ordering, but we cannot choose both. 
Supposing we pick $\left\{ \left| 1 2 \right> , \left| 1 3 \right>, \left| 23 \right> \right\}$, we get
\begin{equation}
\mathbf H_A = \left(
\begin{array}{ccc}
 0 & 1 & 0 \\
 1 & 0 & 1 \\
 0 & 1 & 0
\end{array}
\right), \, \, \, \mathbf H_B = \left(
\begin{array}{ccc}
 0 & 1 & -1 \\
 1 & 0 & 0 \\
 -1 & 0 & 0
\end{array}
\right).
\end{equation}
Forming the evolution operators $\mathbf U = e^{-i t \mathbf H}$, we have
\begin{equation}
\mathbf U_A = \left(
\begin{array}{ccc}
  \left(\frac{\cos \left(\sqrt{2} t\right)}{2}+\frac{1}{2}\right) 
  & -\frac{i \sin \left(\sqrt{2} t\right)}{\sqrt{2}} &  \left(\frac{\cos \left(\sqrt{2} t\right)}{2}- \frac{1}{2}\right) \\
 -\frac{i \sin \left(\sqrt{2} t\right)}{\sqrt{2}} & \cos \left(\sqrt{2}
   t\right) & -\frac{i \sin \left(\sqrt{2} t\right)}{\sqrt{2}} \\
  \left(\frac{\cos \left(\sqrt{2} t\right)}{2} -\frac{1}{2}\right) & -\frac{i \sin
   \left(\sqrt{2} t\right)}{\sqrt{2}} &  \left(\frac{\cos \left(\sqrt{2} t\right)}{2}+\frac{1}{2}\right)
\end{array}
\right)
\end{equation}
and
\begin{equation}
\mathbf U_B = \left(
\begin{array}{ccc}
 \cos \left(\sqrt{2} t\right) & -\frac{i \sin \left(\sqrt{2}
   t\right)}{\sqrt{2}} & \frac{i \sin \left(\sqrt{2} t\right)}{\sqrt{2}}
   \\
 -\frac{i \sin \left(\sqrt{2} t\right)}{\sqrt{2}} & \left(\frac{\cos \left(\sqrt{2} t\right)}{2}+\frac{1}{2}\right) 
 & \left(\frac{1}{2}-\frac{\cos  \left(\sqrt{2} t\right)}{2}\right) \\
 \frac{i \sin \left(\sqrt{2} t\right)}{\sqrt{2}} 
 & \left(\frac{1}{2}-\frac{\cos  \left(\sqrt{2} t\right)}{2}\right)&  \left(\frac{\cos \left(\sqrt{2} t\right)}{2}+\frac{1}{2}\right) 
\end{array}
\right).
\end{equation}
The values of the matrix elements of these two evolution operators clearly have sign differences. 
If we had taken into account that the second and third labels had been switched on graph B, we would have chosen the basis to be $\left\{ \left| 1 2 \right> , \left| 1 3 \right>, \left| 32 \right> \right\}$, fixing the factors of negative one. 
Unfortunately, in a situation where we are handed two graphs and asked whether or not they are isomorphic, we do not know \emph{a priori} what the correct basis choice should be for proper testing. 
Moreover, because the number of possible basis choices is $2^N$, checking all of them is not feasible.
We adopt here a
simple strategy that eliminates this dependence of the sign on the choice of basis
that arises for more than one fermion; we compare
the absolute value of the elements, rather than the elements themselves. 
The absolute values of all the elements and their degeneracies are shown in Table \ref{NIFermiontable}.  The absolute values and degeneracies can be expressed as explicit functions of family parameters, so we conclude that
two non-interacting Fermions, as well as two non-interacting Bosons, fail to distinguish non-isomorphic SRGs from the same family.
The enumeration of the three classes of matrix elements allowed by $\mathbf U_F$ are listed in table \ref{NIFermiontable}, where the $\pm$ symbol indicates that the count given is the total of elements of either sign.

\section{Numerical testing of evolutions of random walks of interacting particles}\label{sec:numerics}
In the preceding two subsections we have proven that
QRWs with two noninteracting particles are not useful for distinguishing non-isomorphic SRGs from the same family.
In this section we perform a systematic investigation of the ability of QRWs of two interacting Bosons to distinguish non-isomorphic SRGs.
We go beyond the sampling performed in \cite{Shiau:2005p1688} by exhaustively checking the two-Boson interacting QRW on all tabulated SRG families with more than one member. 
Our work shows that this walk succeeded in all trials preformed, including successfully distinguishing the (36,15,6,6) SRG family, which has 32,548 non-isomorphic members.
\begin{table}
\caption{Numerical simulations of QRWs of two hard-core Bosons on many SRG families with multiple non-isomorphic members. 
The Hamiltonian used was $\mathbf H_B = -1/2 \cdot \left( \mathbf I + \mathbf S \right) \left( \mathbf A \oplus \mathbf A \right)+ U \mathbf R$, where $\mathbf S$ swaps the two particles, $\mathbf A$ is the adjacency matrix of the graph, and $\mathbf R$ counts double-occupation. To evaluate the hard-core limit, we took $\mathbf U \rightarrow \infty$. 
All non-isomorphic graphs in the families indicated were compared pairwise, with $\Delta$ is a measure of how different the matrix elements of the evolution operator are, defined precisely in eq. \ref{defdelta}. 
When $\Delta = 0$, the list of matrix elements of the two evolution operators being compared have the same magnitudes. 
The minimum values of $\Delta$ were non-zero for all pairs of non-isomorphic graphs that were examined.  
\label{2hcbtable}}
\begin{tabular}{|c|c|c|}
\hline 
SRG family $(N,k,\mu,\lambda)$& non-isomorphic members & minimum $\Delta $ \\
\hline
(16,6,2,0) & 2 & 94.273\\
\hline
(16,9,4,6) & 2 & 2.723\\
\hline
(25,12,5,6) & 15 & 3.636 \\
\hline
(26,10,3,4) & 10 & 7.356 \\
\hline
(28,12,6,4) & 4 & 27.607 \\
\hline 
(29,14,6,7) & 41 & 4.017 \\
\hline
(35,18,9,9) & 227 & 5.243\\
\hline
(36,14,4,6) & 180 & 2.621 \\
\hline
(36,15,6,6) & 32,548 & 1.512 \\
\hline
(37,18,8,9) & 6,760 & 4.310 \\
\hline
(40,12,2,4) & 28 & 3.065\\
\hline
(45,12,3,3) & 78 & 5.868\\
\hline
(64,18,2,6) & 167 & 2.574 \\
\hline
\end{tabular}
\end{table}
We used the following procedure for each pair of graphs in each family:
\begin{enumerate}
	\item Begin with the evolved (complex) evolution matrix $\mathbf U_{A}$.
	\item Take the magnitude of each element.
	\item Write all the (real) entries to a list,  $\mathbf X_{A}$.
	\item Sort the list.
	\item Compare the list using 
	\begin{equation} \label{defdelta}
		\Delta = \sum_{j} \left| \mathbf X_A \left[ j \right] - \mathbf X_B \left[ j \right] \right|.
	\end{equation}
\end{enumerate}
Note that for any two isomorphic graphs, $\Delta=0$, so if $\Delta \neq 0$, we can conclude that the graphs are not isomorphic. 
Table \ref{2hcbtable} presents our results, which show that QRWs of two hard-core Bosons successfully distinguished all pairs of non-isomorphic graphs in all SRG families tested. 

The process of numerically checking that each pair of non-isomorphic graphs was indeed distinguished by our algorithm was computationally intensive.
First, in order to calculate $\mathbf U= e^{-it \mathbf H}$, one must diagonalize $\mathbf H$.
For the $N=64$ cases we considered, since the two-particle space has dimension $64\cdot (64+1)/2 = 2080$, we needed to diagonalize large, non-sparse matrices.
Further, we needed to perform many comparisons to generate all the candidates for the minimum $\Delta$. 
For example, for the (36,15,6,6) family, one needs to perform $32,548 \cdot 32,547 /2 = 529,669,878$ comparisons to check each pairwise $\Delta$.
To accomplish this, we used high-throughput computing environment Condor running on the University of Wisconsin's Center for High Throughput Computing cluster. 
The numerical error on $\Delta$ was between $10^{-14}$ and $10^{-9}$ for all families we analyzed.

In addition to our calculations with hard-core Bosons, we also investigated non-interacting Fermions numerically using the numerical procedure described above.  As discussed above, comparing absolute values of matrix elements eliminates the sign discrepancy brought on by basis choice. 
As one expects given the results in section \ref{secproofF}, the result $\Delta = 0$ is obtained for all cases tested (the first six SRG families appearing in table \ref{2hcbtable}).

\section{Small-time expansion and possible distinguishing operators}\label{sec:smalltime}
In this section we attempt to gain more insight into the behavior of the QRWs of two interacting Bosons by expanding the evolution operator for short times. 
By forming such an expansion and listing all the forms that appear, we can investigate which of these operators contribute to the distinguishing power of the evolution operator.
The operators that contribute are called distinguishing operators, and below we work out the first one, which appears in the fourth order in time. 
We then briefly investigate a sixth-order term, which succeeds in some instances when the fourth-order term fails.

\begin{figure}[tb] 
\begin{center} 
\includegraphics[width=1.0 \linewidth]{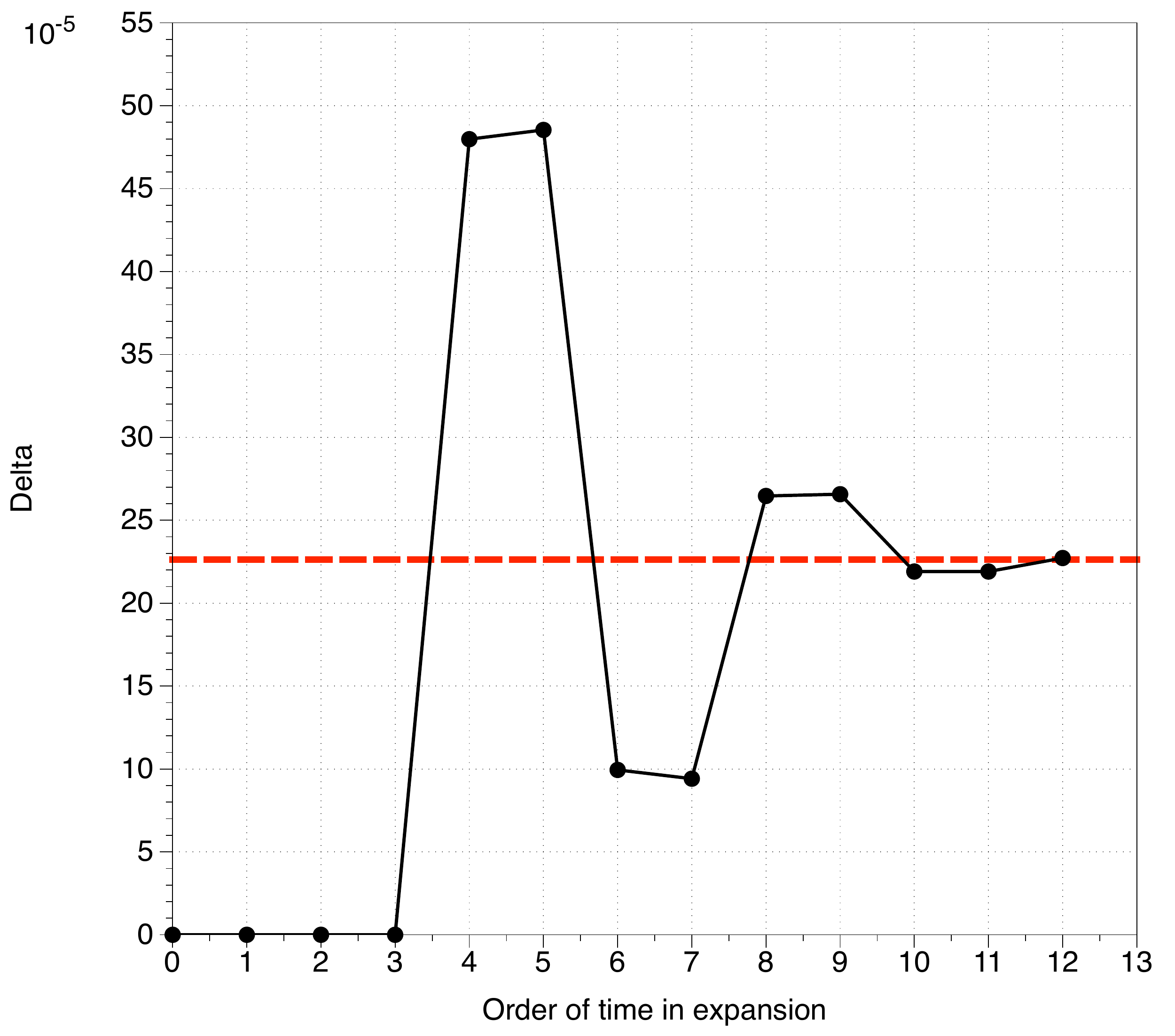}
\end{center} 
\caption{Small-time expansion comparison for the two non-isomorphic SRGs in the family (16,6,0,2) using the interacting two-Boson Hamiltonian $\mathbf H_{2B} = -1/2 \cdot \left( \mathbf I + \mathbf S \right) \left( \mathbf A \oplus \mathbf A \right)+ U \mathbf R$ with $U = 50$, evaluated with $t=0.01$. 
The evolution operator was expanded to the different orders in $t$, and $\Delta$,  a measure of differences in evolution operator matrix elements, is plotted versus the order of the expansion. 
The actual value of $\Delta$ obtained numerically using the full evolution operator is given by the dashed line. 
The distinguishing operator $U \left( \mathbf A \oplus \mathbf A \right) \mathbf R  \left( \mathbf A \oplus \mathbf A \right)^2+U  \left( \mathbf A \oplus \mathbf A \right)^2 \mathbf R  \left( \mathbf A \oplus \mathbf A \right)$, where $\mathbf A$ is the adjacency matrix of the graph, and $\mathbf R$ counts double-occupation, causes the two graphs to be distinguished at fourth order in time.
\label{fig:time_expansion} }
\end{figure}
We begin with the exact evolution operator for the interacting Boson case, which is
\begin{equation}
\mathbf U = e^{-it\left(-\frac{1}{2}\left( \mathbf I + \mathbf S \right) \left( \mathbf A \oplus \mathbf A \right)+ U \mathbf R\right)}.
\end{equation}
Expanding as a power series in $t$, we have
\begin{equation}
\mathbf U = \sum_{n=0}^{\infty}\frac{\left(-i t \right)^n}{n!} \left(-\frac{1}{2}\left( \mathbf I + \mathbf S \right) \left( \mathbf A \oplus \mathbf A \right)+ U \mathbf R\right)^n.
\end{equation}
Since we know from simulation that this evolution operator distinguishes SRGs, we expect that there will be an order in $t$ at which there are terms that are not functions of only the SRG family parameters. 
Numerically, as shown in figure \ref{fig:time_expansion}, we calculate that in the case shown these terms first appear at fourth order in time, so we endeavor to analyze the matrix elements of the first five terms of the expansion for the evolution operator,
\begin{eqnarray}
\mathbf U &\sim& \mathbf U_0 - i t \mathbf U_1  - \frac{t^2}{2} \mathbf U_2 +  \frac{i t ^3}{6} \mathbf U_3 +  \frac{t^4}{24} ~.
\end{eqnarray}
We first expand the terms to simplify the evolution operator, taking $\mathbf A \oplus \mathbf A = \mathbf B$.
\begin{eqnarray}
\mathbf U_1
&=& \frac{1}{2}\left( \mathbf I + \mathbf S \right) \left(U \mathbf R - \mathbf B\right),
\end{eqnarray}
\begin{eqnarray}
\mathbf U_2 
&=& \frac{1}{2}\left( \mathbf I + \mathbf S \right) \left( U^2 \mathbf R+ \mathbf B^2 - U \mathbf R \mathbf B - U \mathbf B \mathbf R \right),
\end{eqnarray}
\begin{eqnarray}
\mathbf U_3 
&=& \frac{1}{2} \left(\mathbf I + \mathbf S \right) \big( U^3 \mathbf R + U \mathbf B^2\mathbf R - U^2 \mathbf B \mathbf R \nonumber \\
&-& U^2 \mathbf R \mathbf B- \mathbf B^3 + U \mathbf R \mathbf B^2 + U \mathbf B \mathbf R \mathbf B \big),
\end{eqnarray}
where we used the fact that any term containing the product $\mathbf R \mathbf B \mathbf R $ vanishes due to the construction of $\mathbf B$. 
Finally,
\begin{eqnarray}
\mathbf U_4
&=&  \frac{1}{2} \left(\mathbf I + \mathbf S \right) \big ( U^4 \mathbf R + U^2 \mathbf B^2\mathbf R - U^3 \mathbf B \mathbf R - U \mathbf B^3 \mathbf R \nonumber \\
&+& U^2 \mathbf R \mathbf B^2 \mathbf R - U^3 \mathbf R \mathbf B - U \mathbf B^2\mathbf R \mathbf B + U^2 \mathbf B \mathbf R \mathbf B  \nonumber \\
&+& U^2 \mathbf R \mathbf B^2 + \mathbf B^4  - U \mathbf R \mathbf B^3 - U \mathbf B \mathbf R \mathbf B^2 \big).
\end{eqnarray}

Numerically, we determined that the fourth order operators that successfully distinguished (16,6,0,2), and thus could not be counted in terms of SRG family parameters, was the combination of $U \mathbf B \mathbf R \mathbf B^2+U \mathbf B^2 \mathbf R \mathbf B$. 
If either of these operators is removed from the fourth order term, $\Delta $ drops to zero. 
Upon further investigation, graphs with more vertices, starting with (25,12,5,6), are not necessarily distinguished by the fourth order in $t$. 
For some, a sixth order expansion was necessary, where we found that at least the term $\mathbf B^2 \mathbf R \mathbf B^3$ helped to distinguish these graphs. 
We tried a sampling of graphs from the families up to $N=40$, and found that this sixth order term succeeded in every instance we tried.

\section{Discussion} \label{sec:discussion}
This paper takes several steps towards characterizing the additional power that quantum walks of interacting particles have for distinguishing non-isomorphic strongly regular graphs, as compared to quantum random walks of noninteracting particles.
We prove analytically that quantum random walks of two noninteracting particles, either bosons or fermions, cannot distinguish non-isomorphic strongly regular graphs from the same family.
We investigate numerically the quantum time-evolution operator for a quantum random walk with two interacting bosons and show
that the resulting Green's functions can be used to distinguish all non-isomorphic pairs of strongly regular
graphs that were investigated.
We perform a much more comprehensive numerical test of the interacting particle algorithm than has been done previously, and find that quantum random walks of two hard-core bosons successfully distinguish all known non-isomorphic pairs of SRGs, which include graphs with up to 64 vertices, and family sizes as large as 32,548 non-isomorphic members.

We now discuss how our results are relevant to possible algorithms for solving GI.
If our algorithm for two hard-core Bosons does indeed distinguish arbitrary graphs, then GI is in P, since the
number of particles is fixed, and only the lattice size increases with number of vertices. 
But, if GI is not in P, then for some pair of graphs, our two particle algorithm must break.
Hence, if such a case was identified, then one could try increasing particle number, which could potentially place GI in BQP,
the complexity class solvable efficiently on a quantum computer.
Unfortunately, we exhausted our test cases (the SRGs), so we could not test this hypothesis.

Although we found that two non-interacting Bosons were not helpful for distinguishing SRGs, we suspect that at larger numbers of non-interacting particles, QRWs might be able to distinguish SRGs. 
This suspicion is due to refs. \cite{mancini-2007-918,Burioni:2000p2289,Buonsante:2002p1723}, where non-interacting QRWs exhibit effective interactions in the statistical limit.
These effective interactions might produce distinguishing power comparable to our explicit hard-core interaction, while providing for an easier analysis.
It would be interesting to examine the several-particle non-interacting QRW to see if this is indeed the case.

\section{Acknowledgements}
This work was supported in part by ARO and DOD (W911NF-09-1-0439).
JKG gratefully acknowledges support from a National Science Foundation Graduate Research Fellowship. 
The authors thank the extremely helpful HEP, Condor, and CHTC groups at Madison for their invaluable assistance in the numerical components of this work.
They also thank Conrad Sanderson for his Armadillo C++ linear algebra library and his help with its use, as well as S-.Y. Shiau for his helpful correspondence on the problem.

\appendix
\section{Checking the two-particle matrix elements}\label{twoptclcheck}
In this appendix we demonstrate that eqs. \ref{bosonoperatorham} and \ref{fermionoperatorham} are equivalent to Eq. \ref{bosonelements}  for both Bosons and Fermions. To show that the Boson matrix elements as given in Eq. \ref{bosonoperatorham} are correct, we evaluate the three types of basis elements we have in eq. \ref{bosonoperatorham}. When $ i \neq j $ and $k \neq l$, Eq. \ref{bosonoperatorham} yields
\begin{eqnarray}\label{bosonop}
\leftsub{B}{\left< i j \right|} \mathbf H_{2B} \left| k l \right>_B 
&=& \left(\frac{ \left< i j \right|' +  \left< j i \right|'}{\sqrt{2}} \right)   \Big[- \frac{1}{2}\left( \mathbf I + \mathbf S \right) \left( \mathbf A \oplus \mathbf A \right) \nonumber \\
&+& U \mathbf R \Big]   \left(\frac{ \left| kl \right>' +  \left| l k \right>'}{\sqrt{2}} \right) \nonumber \\
 &=& - \left(\frac{ \left< i j \right|' +  \left< j i \right|'}{\sqrt{2}} \right) \left( \mathbf A \oplus \mathbf A \right) \\
&\cdot & \left(\frac{ \left| k l \right>' +  \left| l k \right>'}{\sqrt{2}} \right) \nonumber \\
 &=& -A_{ik}\delta_{jl} - A_{jl}\delta_{ik} -A_{il}\delta_{jk} - A_{jk}\delta_{il}. \nonumber
\end{eqnarray}
When $i \ne j$ but $k=l$, we find
\begin{eqnarray}
\leftsub{B}{\left< i j \right|} \mathbf H_{2B}\left| k k \right>_B 
&=& \left(\frac{ \left< i j \right|' +  \left< j i \right|'}{\sqrt{2}} \right) \nonumber \\
&\cdot & \left[- \frac{1}{2}\left( \mathbf I + \mathbf S \right) \left( \mathbf A \oplus \mathbf A \right)+ U \mathbf R \right]  \left| k k \right>' \nonumber \\
 &=& - \left(\frac{ \left< i j \right|' +  \left< j i \right|'}{\sqrt{2}} \right) \left( \mathbf A \oplus \mathbf A \right) \left| k k \right>'  \nonumber \\
 &=& -\frac{1}{\sqrt{2}} \big( A_{ik}\delta_{jk} + A_{jk}\delta_{ik} +A_{ik}\delta_{jk} \nonumber \\
 &+& A_{jk}\delta_{ik} \big) \nonumber \\
  &=& -\frac{2}{\sqrt{2}} \left( A_{ik}\delta_{jk} + A_{jk}\delta_{ik} \right). 
\end{eqnarray}
When $i=j$ and $k=l$, we have
\begin{eqnarray}
\leftsub{B}{\left< i i \right|} \mathbf H_{2B} \left| k k \right>_B 
&=& \left< i i \right| ' \left[- \frac{1}{2}\left( \mathbf I + \mathbf S \right) \left( \mathbf A \oplus \mathbf A \right)+ U \mathbf R \right]  \left| k k \right>' \nonumber \\
 &=& U \left< i i \right| ' \mathbf R \left| k k \right>'  \nonumber \\
 &=& U \delta_{ik}.
\end{eqnarray}
These expressions are all
exactly the same as we found through the definition of the Hamiltonian in Eq.\ ref{bosonelements}. 

We can work the same exercise for the Fermion Hamiltonian, Eq. {fermionoperatorham} as we did for the Boson Hamiltonian in eq. \ref{bosonop}.  The calculation for the fermions yields
\begin{equation}
\leftsub{F}{\left< i j \right|} \mathbf H_{2F}\left| k l \right>_F = \delta_{ik} A_{lj} + \delta_{jl} A_{ik}- \delta_{il} A_{jk} -\delta_{jk} A_{il} ,
\end{equation}
which is identical to the result obtained from eq. \ref{bosonelements}.

\section{Counting the elements in the non-interacting Boson evolution matrix} \label{countingboson}

To prove that the two-particle walk cannot distinguish two non-isomorphic SRGs, we use the method that Shiau et al. introduced for one-particle walks; we show that all the values and degeneracies GFs, matrix elements of the evolution operator, can be expressed of functions of the SRG family parameters. 
For one particle, Shiau et al.~\cite{Shiau:2005p1688} considered the on-diagonal and off-diagonal matrix elements separately. 
For the two-particle evolution operator $\mathbf U_{2B}$, we perform a similar trick by partitioning the matrix elements according to two parameters $(a,b)$: the total number of distinct indices ($a$) and the number of indices shared the left and right sides ($b$). 
For example, $\left< 3 4 \right| U_B \left| 2 4 \right>$ falls into the element class $(3,1)$, since it has three distinct indices (2,3,4) and the left and right side have one index in common (2). 
In total, the two-particle Boson evolution operator has seven such classes: $(4,0)$, $(3,0)$, $(3,1)$, $(2,0)$, $(2,1)$, $(2,2)$, and $(1,2)$, which together partition the set of matrix elements.

Within each of these classes, the various possible element values are listed in table \ref{NIBosontable}. 
Counting the number of occurrences is performed by means of combinatorial sums. 
First, we consider the symmetry class (4,0). 
Since $i \neq j$ and $k \neq l$, we use eq. \ref{defbosonbasis} to write the matrix elements as
\begin{eqnarray}
\left< i j \right| \mathbf U_B \left| kl \right>  &=& \alpha^2 (\delta_{ik} \delta_{jl} + \delta_{il} \delta_{jk}) + 2 \beta^2 \nonumber \\
&+& \gamma^2(A_{ik} A_{jl} + A_{il} A_{jk}) \nonumber \\
&+& \alpha \beta (\delta_{ik} + \delta_{jk} + \delta_{il} + \delta_{jl} ) \nonumber \\
&+& \alpha \gamma (\delta_{ik} A_{jl} + \delta_{jk} A_{il} + \delta_{il} A_{jk} + \delta_{jl} A_{ik} ) \nonumber \\
&+& \beta \gamma (A_{jl} + A_{il} + A_{jk} + A_{ik}).
\end{eqnarray}
One possible value for this matrix element, $4 \beta \gamma + 2 \gamma^2+2\beta^2$, occurs when $A_{jl} = A_{il} = A_{jk} = A_{ik} =1$. 
Since $\mathbf A$ is a $0-1$ matrix, the number of ways this can occur, $n_{(4,0)a}$, is given by
\begin{eqnarray}
n_{(4,0)a} &=& \sum_{i<j} \sum_{k<l}A_{jl}  A_{il}  A_{jk}  A_{ik} \\
&=& \frac{1}{4}\sum_{ijkl} A_{jl}  A_{il}  A_{jk}  A_{ik} (1 - \delta_{ij}) ( 1 - \delta_{kl}) \nonumber \\
&=& \frac{1}{4} \Big( \sum_i (A^4)_{ii} - 2 \sum_{ij} (A^2)_{ij} + \sum_ij A_ij \Big), \nonumber
\end{eqnarray}
where the initial sum is constrained to $i<j$ and $k<l$ since we are working with indistinguishable Bosons, and hence have a space of dimension $N(N+1)/2$. 
By repeated use of eq. \ref{SRGAlgebra}, we can use the techniques of the one-particle algorithm~\cite{Shiau:2005p1688} to evaluate these sums in terms of family parameters. 
The values of those pertinent to our present discussion are:
\begin{eqnarray} \label{sumrules}
\sum_{ij}A_{ij} &=& k N \nonumber \\
\sum_{ij}(A^2)_{ij} &=& N (k - \mu) +k N (\lambda - \mu ) + N^2 \mu \nonumber \\
\sum_{ij}(A^3)_{ij} &=& N \left(k^2+k \big(\mu  (N+\mu -2)+\lambda ^2-2 \lambda  \mu +\lambda \right) \nonumber \\
&+&(N-1) \mu  (\lambda -\mu )\big) \nonumber \\
\sum_{i}(A^3)_{ii} &=& k N \lambda \nonumber \\
\sum_{i}(A^4)_{ii} &=& k N \left(\mu  (k-\lambda -1)+k+\lambda ^2\right).
\end{eqnarray}
Plugging in these results and using the SRG relationship $(N-k-1) \mu = k (k - \lambda - 1)$, we find
\begin{eqnarray}
n_{(4,0)a} &=& 1/4 \cdot N \big(k^2 (\mu +1)+k \left(\lambda ^2-\lambda  (\mu +2)+\mu -1\right) \nonumber \\
&-& 2 (N-1) \mu \big).
\end{eqnarray}
A second possible value for this matrix element is $3 \beta \gamma + \gamma^2 + 2 \beta ^2$, obtained by setting any one of $A_{jl}$,  $A_{il}$, $A_{jk}$, or $A_{ik}$ to zero. 
As a sum, this means that the number of occurrences, $n_{(4,0)b}$, is
\begin{eqnarray}
n_{(4,0)b} &=& 4  \sum_{i<j} \sum_{k<l}A_{jl}  A_{il}  A_{jk} (1- A_{ik})(1-\delta_{ik}) \nonumber \\
&=&\sum_{ijkl} A_{jl}  A_{il}  A_{jk} (1- A_{ik})(1-\delta_{ik})(1 - \delta_{ij}) ( 1 - \delta_{kl}) \nonumber \\
&=& \sum_{ij }(A^3)_{ij} - \sum_i (A^3)_{ii} - \sum_i(A^4)_{ii} \nonumber \\
&=& N \mu  (N-k-1) (k+\lambda -\mu ),
\end{eqnarray}
where the initial factor of four is due to the four possible ways to pick the $A$, and $ (1- A_{ik})(1-\delta_{ik})$ constrains both $i \neq k$ and $A_{ik} = 0$. 
The remainder of the calculation proceeds similarly, and the results are listed in table \ref{NIBosontable}. 

\bibliography{two-particles}

\end{document}